\documentclass[conference]{IEEEtran}
\IEEEoverridecommandlockouts
\usepackage{cite}
\usepackage{amsmath,amssymb,amsfonts}
\usepackage{algorithmic}
\usepackage{graphicx}
\usepackage{textcomp}
\usepackage{xcolor}
\def\BibTeX{{\rm B\kern-.05em{\sc i\kern-.025em b}\kern-.08em
    T\kern-.1667em\lower.7ex\hbox{E}\kern-.125emX}}

\usepackage{xcolor,colortbl}
\usepackage{amsthm}
\usepackage{listings}
\usepackage{ifthen}
\usepackage{color}
\usepackage{booktabs}
\usepackage{url}

\newcommand\VB{{VulBERTa}}
\newcommand\cmd[1]{{\small \texttt{#1}}}
\newcommand\scmd[1]{{\scriptsize \texttt{#1}}}
\newcommand\VBM{{VulBERTa-MLP}}
\newcommand\VBC{{VulBERTa-CNN}}

\begin{document}

\title{VulBERTa: Simplified Source Code Pre-Training for Vulnerability Detection \thanks{\textbf{Accepted as a conference paper at IJCNN 2022.}}
}

\author{\IEEEauthorblockN{Hazim Hanif}
\IEEEauthorblockA{\textit{Department of Computing} \\
\textit{Imperial College London, UK;}\\
\textit{Faculty of Computer Science and Information Technology,}\\
\textit{University of Malaya, Malaysia}\\
m.md-hanif19@imperial.ac.uk}
\and
\IEEEauthorblockN{Sergio Maffeis}
\IEEEauthorblockA{\textit{Department of Computing} \\
\textit{Imperial College London, UK}\\
sergio.maffeis@imperial.ac.uk}
}

\maketitle

\begin{abstract}
    This paper presents VulBERTa, a deep learning approach to detect security vulnerabilities in source code. Our approach pre-trains a RoBERTa model with a custom tokenisation pipeline on real-world code from open-source C/C++ projects. The model learns a deep knowledge representation of the code syntax and semantics, which we leverage to train vulnerability detection classifiers. We evaluate our approach on binary and multi-class vulnerability detection tasks across several datasets (Vuldeepecker, Draper, REVEAL and muVuldeepecker) and benchmarks (CodeXGLUE and D2A). The evaluation results show that VulBERTa achieves state-of-the-art performance and outperforms existing approaches across different datasets, despite its conceptual simplicity, and limited cost in terms of size of training data and number of model parameters.
\end{abstract}


\section{Introduction}

%

%
MITRE \cite{MITRE2021cvedetails} reports an increase in the number of software CVEs submitted yearly since 2016, reflecting the increased threat to the overall security of the software ecosystem. Accordingly, there has been a steady growth of research in software vulnerability detection over the years \cite{Hanif2021Rise}, across the spectrum of vulnerability detection approaches such as static analysis, dynamic analysis and machine learning-based detection models.

Deep learning has obtained encouraging results for software vulnerability, in particular using sequence- and graph-based techniques such as Bidirectional-LSTM \cite{li2018vuldeepecker,Li2021SySeVR,zou2019muvuldeepecker} and Graph Neural Networks \cite{Zhou2019Devign,cao2021bgnn4vd}. 
These techniques attempt to embed syntactic and semantic information from the code explicitly, for example by using various dependency and data flow analyses to preprocess source code and extract various artefact such as code gadgets, control flow graphs and dependency graphs which are eventually fed to the respective neural network. 

In this paper, we follow a different approach inspired by the recent successes of Transformer-based neural architectures~\cite{vaswani2017tranformers,devlin-etal-2019-bert, lan2020albert, he2021deberta, wang-etal-2018-glue} able  to learn deep representation knowledge of natural language textual data.
Our aim is to build a representation model of C/C++ that embeds syntactic and semantic information about the language without our direct intervention, and then use that  model as a basis to build vulnerability detection models using standard neural architectures.

We present \VB, which aims to learn a deep representation of C/C++ source code from large codebases that consist of different software projects. To facilitate learning the internal code representation, we need to create a language-aware reliable tokenisation pipeline, to parse and tokenise source code while ensuring that basic, yet key syntactic and semantic information is made available to the model. In our tokenisation pipeline, we implemented several novel features to enhance the tokenisation ability of the existing Byte-pair Encoding (BPE) \cite{gage1994bpe} tokeniser. We introduce a custom tokeniser that combines BPE with custom pre-defined code tokens to build the vocabulary of our tokeniser. These pre-defined tokens are based on the AST node type of Clang \cite{LLVM2021libclang} (standard keywords and punctuation) and lists of common C/C++ API function names. Our custom tokenisation pipeline creates a better code representation while maintaining its original syntactic structure even after being encoded.

\VB\ implements a Transformer-based deep learning architecture called RoBERTa \cite{liu2019roberta}. \VB\ builds its code representation knowledge via Masked Language Modelling (MLM) \cite{devlin-etal-2019-bert}. We pre-train \VB\ on fewer samples (2.28 million) and model parameters (125 million) than existing approaches \cite{phan2021cotext,buratti2020cbert}. Then, we connect the pre-trained \VB\ model with alternatively a MultiLayer Perceptron (\VBM)\ and a Convolutional Neural Network (\VBC) in order to fine-tune vulnerability detection models.

We evaluate \VB\ across different datasets to test the performance and transferability of our pre-trained model. The evaluation results show that we achieve state-of-the-art detection performance across these datasets. In particular, we achieved an F1 score of 57.92\% on the Draper \cite{Russell2018Draper} dataset, outperforming existing approaches, which is encouraging as this dataset is considered challenging due to being imbalanced and having a mix of real-world and synthetic samples. \VB\ also performs well on multi-class classification task on the easier muVuldeepecker \cite{zou2019muvuldeepecker} dataset, with a weighted F1 score of 99.59\%. 
We also tested \VB\ on two software vulnerability detection benchmarks (CodeXGLUE and D2A), where it  outperforms most existing approaches, despite using less training data and fewer model parameters.
These results show that \VB\ is effective in learning a deep representation of source code, and using that knowledge to detect software vulnerabilities.

In summary, our main contributions are: 
\begin{itemize}
    \item A custom tokenisation pipeline which combines the BPE algorithm with novel pre-defined code tokens (standard C/C++ keywords, punctuation and library API calls), providing better code encodings while maintaining the syntactical structure of source code.
    \item A small and simplified pre-trained model, (\VB)\ that provides transferability of pre-trained embedding weights, which is reusable in less complex architectures such as MLP and CNN.
    \item Software vulnerability detection models, \VBM\ and \VBC\ that achieve state-of-the-art (SOTA) detection performance across different datasets and top-3 positions in two benchmarks (CodeXGLUE and D2A).
\end{itemize}

Our code and data are open source, and made available at \textit{\url{https://github.com/ICL-ml4csec/VulBERTa}}.

\section{Related Work}
This Section discusses related work on software vulnerability detection using deep learning, and more in general on pre-training models for representing programming languages.

\subsection{Vulnerability detection using deep learning}

Software vulnerability detection remains a challenging problem for security researchers across academia and industry. 
Much work has been done to reliably and efficiently detect security vulnerabilities in source code \cite{6956589joern,Henning2015vccfinder,stuckman2017effect}. However, recent advances in deep learning and its application across different domains have induced security researchers to test the effectiveness of deep learning on vulnerability detection tasks. 

One of the earliest works that use deep learning techniques to detect software vulnerabilities on raw source code is~\cite{Russell2018Draper}. The unusual aspect of this work is that instead of using CNN and RNN to train a classifier model, they used it as a feature extractor. The extracted features are then passed to a Random Forest (RF) classifier trained to detect software vulnerabilities. This approach achieves an AUC score of 90.4\% when tested on real-world dataset.

Vuldeepecker~\cite{li2018vuldeepecker} proposes a structure called code gadgets that is based on the extraction of library/API function calls from the source code. 
However, a limitation of this work is that it only considers vulnerability that involves library/API function calls. To overcome such limitations, the authors extended their work and proposed SySeVR \cite{Li2021SySeVR} as a systematic framework for detecting vulnerabilities and can learn from syntax and semantic of the source code. They updated the code gadgets approach to accommodate both data and control dependency information from the code. 
Following the success of code gadgets in Vuldeepecker and SySeVR, muVuldeepecker \cite{zou2019muvuldeepecker} was introduced to detect different types of vulnerabilities through multi-class classification task. In addition, this work  proposed a mechanism to pinpoint the location of a specific vulnerability in the source code. 

Independently from the work on code gadgets, other research investigated graph-based approaches for vulnerability detection. 
Devign \cite{Zhou2019Devign} implemented a Graph Neural Network (GNN) model to learn data and control dependency code graphs and proposed a novel Conv module that extracts interesting features from  source code. DeepWukong \cite{Chenf2021Deepwukong} embeds code fragments in a compact and low-dimensional representation to detect ten different types of vulnerabilities using GNN. 
%
Beyond C/C++, DeepTective \cite{rabheru2022deeptective} detects SQLi, XSS and command injection vulnerabilites in PHP source code by combining Gated Recurrent Units (GRU) and Graph Convolutional Networks.
Hybrid neural networks were also explored in \cite{YAN201867}, where the authors constructed a Hybrid Deep Learning Network to detect code injection attacks in HTML5-based applications.

Another interesting line of research for deep learning-based vulnerability detection focuses on dataset quality. REVEAL~\cite{chakraborty2020reveal} leverages the SMOTE re-sampling and duplicate removal method to address the problem of imbalanced datasets. 
D2A~\cite{9402126d2a} proposes a curated benchmark dataset based on a differential analysis approach, by analysing version pairs of source code from multiple open-source projects.

\subsection{Pre-trained models of source code}

Pre-training is a technique used by the Transformer~\cite{vaswani2017tranformers} neural architecture, initially introduced in the Natural Language Processing (NLP) domain to learn deep representation knowledge of textual data, for  language-specific tasks such as text translation, text completion and text generation \cite{devlin-etal-2019-bert, lan2020albert, he2021deberta, wang-etal-2018-glue}. 
%

To investigate and test the hypothesis of whether programming languages can be understood using NLP techniques, \cite{buratti2020cbert} presented C-BERT, a pre-training architecture that is based on Abstract Syntax Trees (AST). The main idea behind this is to learn AST-based features automatically from source code during pre-training. 
In the fine-tuning task, C-BERT outperformed existing approaches across AST node tagging and vulnerability detection tasks. 

CodeBERT \cite{feng-etal-2020-codebert} implemented BERT \cite{devlin-etal-2019-bert}, one of the earliest pre-training architectures, on programming languages. The authors pre-trained the model using bimodal and unimodal data, consisting of code and natural language pairs from different programming languages such as Java and Python. 
CodeBERT achieved high BiLingual Evaluation Understudy (BLEU) score as compared to RoBERTa \cite{liu2019roberta} models.

DOBF~\cite{roziere2021dobf} proposes a new pre-training objective that is based on code deobfuscation. The deobfuscation objective focuses on the structural aspect of programming language, whether pre-training helps the model to learn syntactic and structural information of the source code.
Besides that, in order to learn and investigate large graph-based representation, GraphCodeBERT~\cite{guo2021graphcodebert} proposed a pre-trained model that incorporates graph structure into the Transformer-based model using graph-guided masked attention to filter irrelevant signals. In the evaluation, GraphCodeBERT showed a strong performance compared to other pre-trained models (RoBERTa and CodeBERT) in different tasks such as code clone detection, code translation, and code refinement.

\begin{figure*}[!t]
  \scriptsize
 \centering
 \includegraphics[scale=0.75]{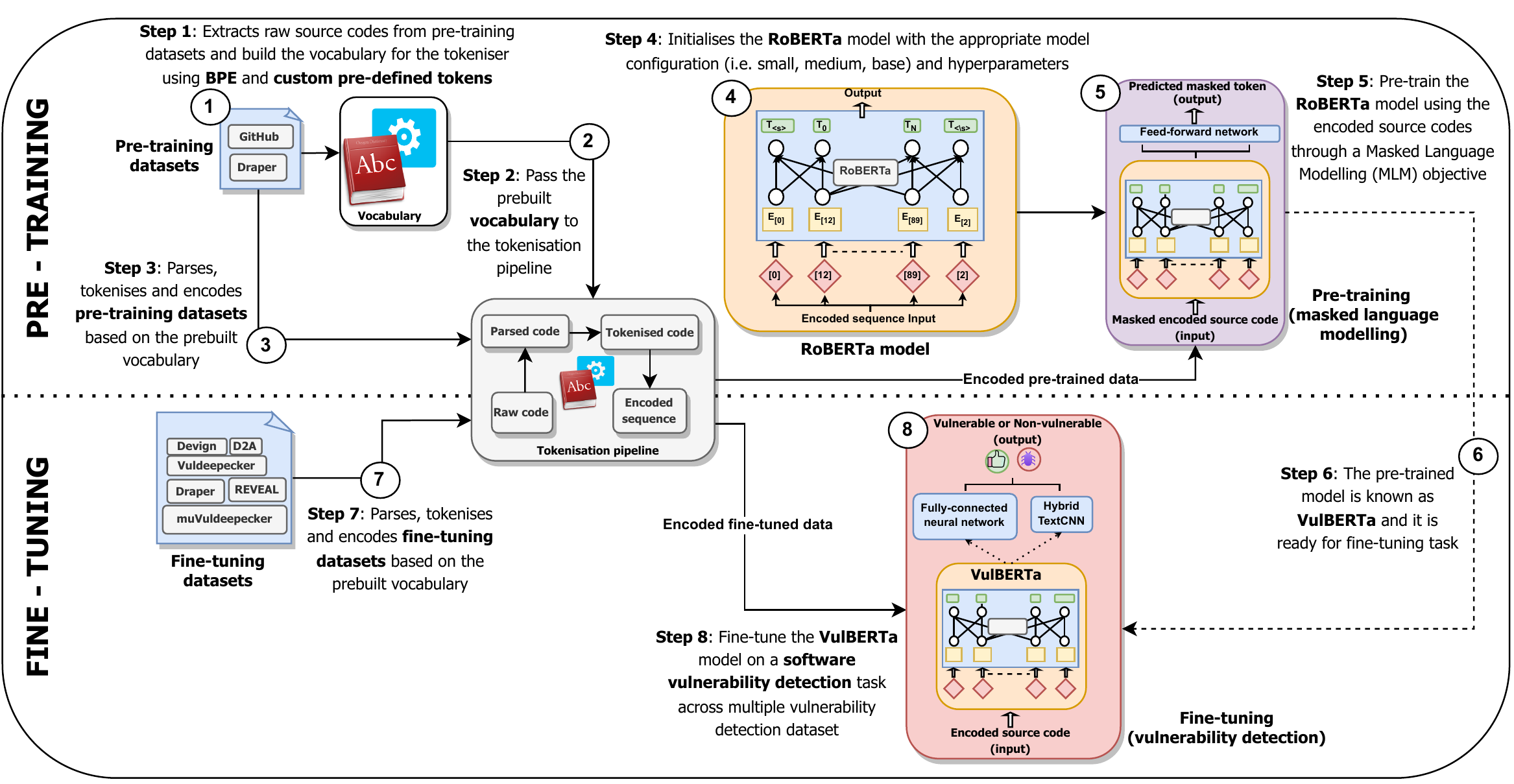}\vspace{-5pt}
    \caption{\VB\ training pipeline. Steps are taken in order from 1 to 8.}\label{fig:vb_arch}
\end{figure*}

\section{\VB}
In this section, we introduce VulBERTa, our pre-training architecture for detecting vulnerabilities in C/C++ source code at function-level granularity. The architecture is divided into three key components: a tokenisation technique that parses and tokenises code using a custom vocabulary; a pre-training session that builds a representation model for code; and a fine-tuning session that refines the model to target a concrete classification task. Figure \ref{fig:vb_arch} visualises the 8 main steps of the \VB\ training pipeline.

\subsection{Tokeniser}

\begin{figure}[t]
  \scriptsize
 \centering
 \includegraphics[scale=0.55]{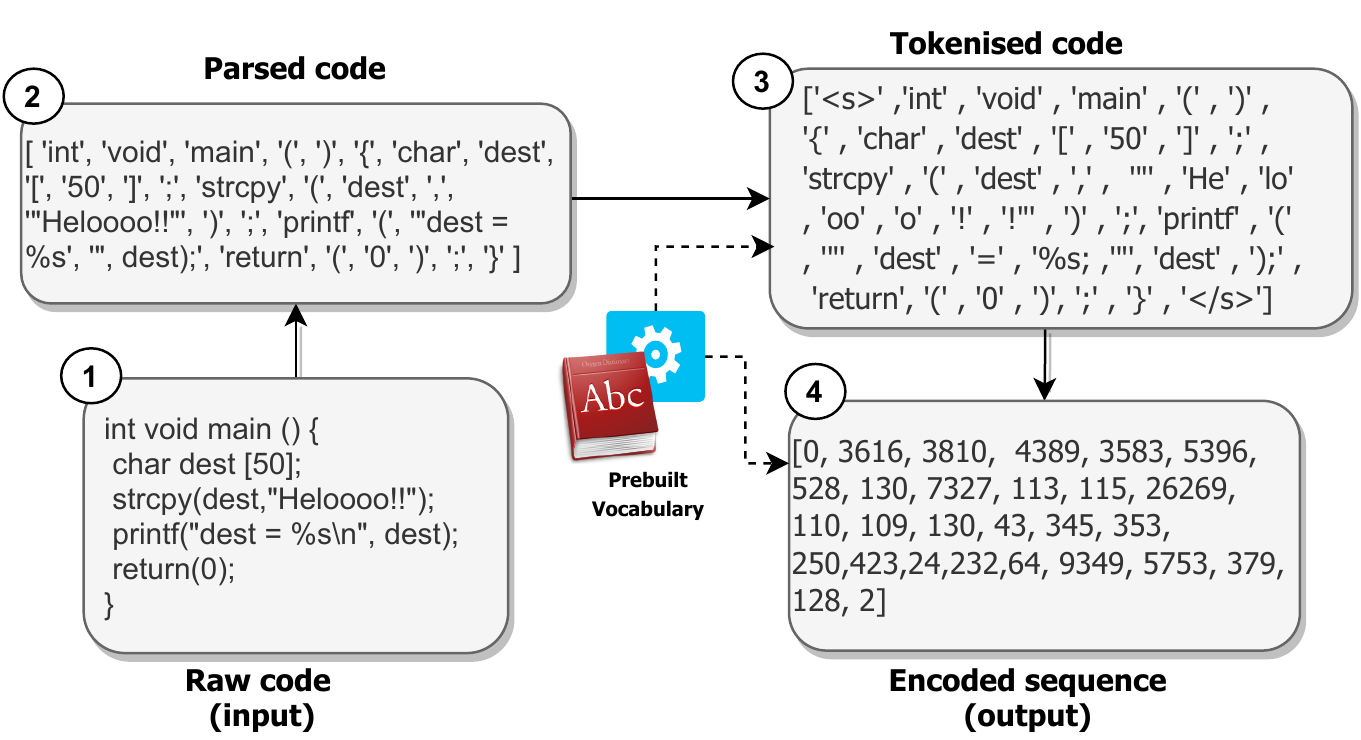}\vspace{-5pt}
    \caption{Tokenisation pipeline.}\label{fig:tokenisation_pipeline}
\end{figure}

Our tokenisation pipeline aims at preserving syntactic structure and selected semantic identifiers. It consists of a parser, a tokeniser and an encoder. These components stack onto each other to transform raw source code into a structure understandable by a neural network. Figure \ref{fig:tokenisation_pipeline} visualises the overall tokenisation pipeline of \VB\ from raw code to encoded output sequence. Below, we describe each step of the pipeline.

\subsubsection{Parser}
We remove comments from the source code of each function using several regular expressions. 
Then, we parse the source code using Clang \cite{LLVM2021libclang}, a robust C/C++ parser that can parse code without including any libraries or external dependencies. 
Clang allows us to preserve the syntactic structure of the source code while breaking it down into a sequence of code tokens. 

\subsubsection{Tokenisation}
The tokens produced by the Clang parser are further processed by the BPE algorithm, modified to take into account our pre-defined tokens, to further break the parsed input down into fine-grained tokens for encoding.

\textbf{Byte Pair Encoding (BPE)} is a subword tokenisation algorithm proposed by \cite{gage1994bpe} that replaces a similar pair of consecutive bytes with a byte that does not appear in the data.
Subword tokenisation also reduces the possibility of encountering out-of-vocabulary tokens as most subword tokens are available in the vocabulary. Following several implementations of BPE, such as in \cite{buratti2020cbert,feng-etal-2020-codebert}, we define a vocabulary size of 50000 as our maximum number of entries in the vocabulary.

\textbf{Pre-defined tokens} are tokens we explicitly include in the vocabulary, thus excluding them from the subword tokenisation process. Our goal is to preserve their syntactic or semantic meaning. We have considered using token bucketing, normalization and standard C/C++ tokens. We found that by pre-defining C/C++ keywords, punctuation, and standard API names, we preserve more information about the meaning of the source code during pre-training.
Table \ref{tab:predef_token} summarises our pre-defined tokens, which are excluded from BPE. The full list consists of 451 pre-defined tokens. The other tokens consist of literals and identifiers, and are passed to BPE to undergo the tokenisation process. 

\begin{table}[t]
    \scriptsize
    \caption{Custom pre-defined tokens.}\vspace{-5pt}
    \label{tab:predef_token}
    \centering
    \begin{tabular}{l l l}
    \toprule
    \textbf{Token type} & \textbf{Total} & \textbf{Examples}\\
    \midrule
    BPE reserved tokens & 5 & \scmd{<pad>}, \scmd{<unk>}, \scmd{<mask>}\\
    Standard C/C++ keyword tokens & 104 & \scmd{int}, \scmd{if}, \scmd{void}\\
    Standard C/C++ punctuation tokens & 54 & \scmd{=}, \scmd{++}, \scmd{->}\\
    Standard C/C++ API call tokens & 288 & \scmd{strlen}, \scmd{scanf}, \scmd{memcpy}\\
    \bottomrule
    \end{tabular}
\end{table}

\subsubsection{Encoder}
Encoding is the process of converting the code tokens into tensors. 
For pre-training, we have set the maximum sequence length to 512 to maximise and generalise the learning throughout the data, as the pre-training dataset comprises different codebases. 
Meanwhile, we increase the maximum sequence length to 1024 for the fine-tuning task so that it can contain, without truncation, more than 90\% of the actual samples on average. We pad shorter sequences to the right, with a special padding token (\cmd{<pad>}).

\subsection{Pre-training}
Pre-training is the initial training session where we train a standard RoBERTa \cite{liu2019roberta} model with MLM objective in order to learn an informative general representation of C/C++ code across different software project.
We will refer to the model obtained from this pre-training session as the \textbf{\VB\ model}.
Combining different software projects is beneficial to the learning process, as it generalises the representation knowledge of the code even further across different coding styles. This also helps to increase the robustness of the model during pre-training.
We set the embedding size to be 768 dimensions, following RoBERTa-base.
This is the core embedded knowledge of the model, that will be useful for downstream fine-tuning tasks. 
%
%

\subsection{Fine-tuning}
\begin{figure}[t]
  \scriptsize
 \centering
 \includegraphics[scale=0.5]{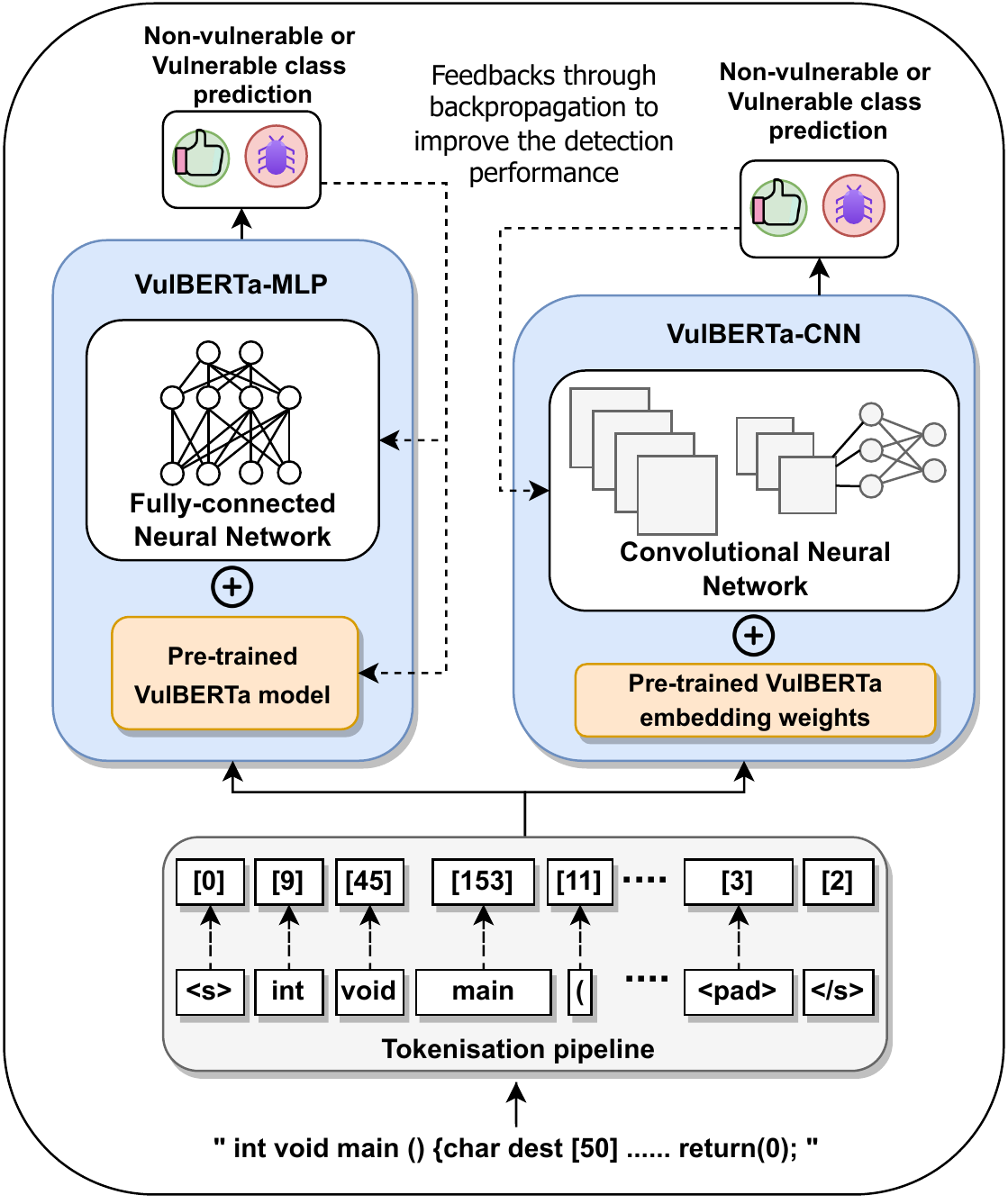}\vspace{-5pt}
    \caption{Fine-tuning (vulnerability detection) pipeline.}\label{fig:fine_tune}
\end{figure}

In fine-tuning, we further train the pre-trained model on a specific downstream task, which in our case is software vulnerability detection.
%
%
Figure \ref{fig:fine_tune} shows our fine-tuning pipeline for vulnerability detection.
%
We implement two different classification approaches for fine-tuning. The first approach is a standard multilayer perceptron (MLP) on top of the pre-trained model, and the second approach uses a Text Convolutional Neural Network (TextCNN), which is cheaper and faster to fine tune due to the robustness of CNN architecture.

\textbf{\VBM:}
We implement a fully-connected layer with 768 neurons and one output layer 2 or 41 neurons based on whether our fine-tuning dataset is a binary or multi-class classification dataset. During fine-tuning, we reuse the pre-trained weights from \VB\ and continue the training for several epochs. This approach is the most common approach for fine-tuning pre-trained models as it takes advantage of the whole VulBERTa architecture with little modification \cite{liu2019roberta,devlin-etal-2019-bert}.

\textbf{\VBC:}
We extract the embedding weights of the pre-trained \VB\ model and use them as the embedding weights for a Hybrid  TextCNN~\cite{kim2014textcnn}. 
The TextCNN architecture consists of three 1-dimensional CNN, each with its max-pooling layer. The outputs are concatenated and flattened, and then fed to two fully-connected layers (256 and 128 neurons) with one output layer for classification. 
Since the embeddings are already pre-trained on a large language model, we freeze them during the training task. This technique allows the TextCNN model to inherit and use the representation knowledge of the embeddings and focus on tuning the weights of the task-specific CNN layers.

\section{Datasets}\label{sec:datasets}
This Section describes the datasets used in the rest of the paper. They consist of function-level C/C++ source code from various codebases, including open-source repositories and synthetic code samples. We divide these datasets in two categories based on their prevalent use for either pre-training or fine-tuning. All datasets mentioned below are in the public domain, and available for download.

\subsection{Pre-training datasets}
For pre-training, we use the Masked Language Modelling (MLM) task on the GitHub and Draper datasets.


\subsubsection{GitHub}
The GitHub dataset consists of the source code of 1,101,075 C/C++ functions extracted from 1060 open-source repositories on GitHub. We compiled this dataset using the public GitHub API and the \cmd{PyGithub}~\cite{PyGithub2021pygithub} package. First we gathered all the names of repositories containing C/C++ source code, together with their stars count. Then we sorted these repositories based on their star counts, and started fetching files from the repositories. This process took around two days to complete due to API rate-limit restrictions. Finally  we used Joern~\cite{6956589joern}, an open-source code querying engine for C/C++, to efficiently extract individual functions from each downloaded file. This step is essential as our approach aims to detect function-level security vulnerabilities.

\subsubsection{Draper}\label{lab_draper}
The Draper dataset is a software vulnerability detection dataset initially introduced by \cite{Russell2018Draper}. This dataset consists of 1,274,366 C/C++ functions gathered from various locations such as Debian Linux distributions, open-source GitHub repositories and the Juliet test suite~\cite{NIST2017juliet}. This dataset ranges from highly documented production code to synthetic test samples.


\subsection{Fine-tuning datasets}\label{finetune_datasets}
For the fine-tuning task, which in our case is vulnerability detection, we have selected a number of
datasets which security researchers have compiled in order to evaluate their respective vulnerability detection approaches. 

\subsubsection{Vuldeepecker}
The Vuldeepecker dataset is a vulnerability detection dataset introduced in \cite{li2018vuldeepecker}. It consists of real-world samples from the National Vulnerability Database (NVD) \cite{nvd} and synthetic samples from the Software Assurance Reference Dataset (SARD) \cite{sard} project. These two projects are actively maintained by the National Institute of Standards and Technology (NIST). This dataset is frequently used as a benchmark dataset to evaluate vulnerability detection techniques for C/C++ source code. 

\subsubsection{Draper}
This is the same dataset we described in Section \ref{lab_draper}, however for fine-tuning we include the (binary) labels. This dataset was checked by three static analysers and labelled by a team of security experts.

\subsubsection{REVEAL}
The REVEAL dataset is a real-world software vulnerability detection dataset introduced in \cite{chakraborty2020reveal} in response to existing datasets that contain lots of data duplication and unrealistic distribution of vulnerable classes. This dataset is a binary detection dataset consisting of source code from two open-source projects: Linux Debian kernel and Chromium. 

\subsubsection{muVuldeepecker (MVD)}
The muVuldeepecker dataset is a multiclass vulnerability detection dataset introduced in \cite{zou2019muvuldeepecker}. It is remarkably similar to the Vuldeepecker dataset as this dataset also comes from the NVD and SARD. However, the main difference is that this dataset consists of code gadgets instead of the usual function-level source code.

\subsubsection{Devign} \label{dataset:devign}
The Devign dataset is a real-world software vulnerability detection dataset initially introduced in \cite{Zhou2019Devign}. This dataset consists of function-level C/C++ source code from two popular open-source software projects, QEMU and FFmpeg. The labelling and verification have been done manually by a team of security researchers over a two-round process.

\subsubsection{D2A} \label{dataset:d2a}
The D2A dataset is a real-world vulnerability detection dataset curated and introduced by the IBM Research team \cite{9402126d2a}. This dataset consists of several open-source software projects like FFmpeg, httpd, Libav, LibTIFF, Nginx and OpenSSL. It was created using a differential analysis technique to label issues reported by static analysers.

\section{Software vulnerability detection}\label{sec:vulns}
In this Section we describe how we pre-train the \VB\ model, build fine-tuned \VBM\ and \VBC\ models, and evaluate them across several vulnerability detection datasets and benchmarks. 

\subsection{Experimental setup}

\subsubsection{Hardware and software}
We use PyTorch 1.7 \cite{paszke2019pytorch} with CUDA 10.2 on top of Python 3.7 for all fine-tuning experiments.  For pre-training we use Google Compute Engine (GCP) VMs that have 48 vCPUs, 240GB RAM and 2 NVIDIA Tesla A100 40GB GPUs. For fine-tuning, we use a machine with 48 cores Intel Xeon Silver CPU, 292GB RAM and 2 NVIDIA GTX TITAN Xp GPU. Each GPU has 12GB of video memory to accommodate different model configurations.

\subsubsection{Performance criteria}
For every experiment, we report several evaluation metrics, including those used in the initial work for each dataset. This way, we can perform a fairer comparison. The metrics are true negatives (TN), false negatives (FN), true positives (TP), false positives (FP), accuracy, precision, recall, F1-score, Receiver Operating Characteristic area under the curve (ROC-AUC), precision-recall AUC (PR-AUC), and Matthews Correlation Coefficient (MCC).

\subsubsection{Baselines methods}
In the performance evaluation, we compare \VB\ with two baselines techniques on top of existing approaches for each dataset. These two baselines are widely used to analyse sequence-based input for vulnerability detection and have been used for example in \cite{Zhou2019Devign} and \cite{buratti2020cbert}.

\textbf{(i) Baseline-BiLSTM:} This technique is a variation of LSTM, which implements a two layers Bi-directional LSTM \cite{GRAVES2005bilstm} and several fully-connected layers to learn from sequences of source code for vulnerability detection. The bi-directional LSTM learns forward and backward relationships of code sequence simultaneously.

\textbf{(ii) Baseline-TextCNN:} This is a variant of CNN \cite{kim2014textcnn}, where the input data is natural language text instead of images. In this case, we use source code as the input data and feed it into CNN. This technique deploys three convolutional layers with pooling, and concatenates them into a single layer before passing the results to several fully-connected layers.

\subsubsection{Model pre-training}\label{model_pretraining}
We pre-train the \VB\ model with Draper and GitHub datasets using MLM. We experiment with different RoBERTa configurations (i.e. \emph{small}, \emph{medium} and \emph{base}) to see how the number of model parameters affects the pre-training performance of the model.
%
The duration of each pre-training session falls between 72 to 96 hours depending on the model configuration. The training session is done up to 500,000 steps and a learning rate scheduler is used to reduce the learning rate over time as the training loss plateaus. 
Based on the results, we found that the Base model gives the lowest loss during training after 500,000 steps. Therefore, we choose the \VB\ model with Base configuration ($\sim$125M parameters) as our reference pre-trained model for fine-tuning.


\subsubsection{Model fine-tuning}
We fine-tune our pre-trained \VBM\ and \VBC\ models separately, on each dataset described in Section~\ref{finetune_datasets}, using vulnerability detection as the fine-tuning objective. 
 We set the maximum number of epochs to 10, which is more than sufficient as the models tend to start overfitting to the training set after 4-5 epochs. We set the learning rate to 0.00003 with a learning rate scheduler to reduce the learning rate as the training loss plateaus. We follow the original split for each dataset, but if the split information is unavailable, we split the dataset 80/10/10 (training/validation/testing). 
 Each fine-tuning session lasted between 5 and 10 hours, depending on the size of the dataset and model.

\subsection{Evaluation on selected datasets}

The vulnerability detection experiments are separated by dataset. 
%
%
%
For each dataset, we selected the preferred evaluation metric (PEM) used for comparison in the original paper. Table~\ref{tab:finetuning_results} shows the evaluation results, where we highlighted the highest PEM score for each dataset.

\begin{table*}[t]
    \scriptsize
    \caption{Dataset evaluation results: summary of evaluation metrics}\vspace{-5pt}
    \label{tab:finetuning_results}
    \centering
    \begin{tabular}{p{1.5cm} p{0.9cm} | p{2cm} p{0.3cm} p{0.4cm} p{0.5cm} p{0.4cm} p{0.3cm} p{0.5cm} p{0.3cm} p{0.5cm} p{1cm} p{1.2cm} p{0.5cm} p{1cm}}
    \toprule
    \textbf{Dataset} & \textbf{PEM} & \textbf{Model} & \textbf{FN} & \textbf{FP} & \textbf{TN} & \textbf{TP} & \textbf{Acc (\%)} & \textbf{Prec (\%)} & \textbf{F1 (\%)} & \textbf{Recall (\%)} & \textbf{PR-AUC (\%)} & \textbf{ROC-AUC (\%)} & \textbf{MCC (\%)} & \textbf{Weighted F1 (\%)}\\
    \midrule
    Vuldeepecker & Precision & Vuldeepecker \cite{li2018vuldeepecker}& - & - & - & - & - & 91.90 & 92.90 & - & - & - & - & - \\
     &  & \VBM\ & 99 & 39 & 15002 & 881 & - & \textbf{95.76} & 93.03 & - & - & - & - & - \\
     &  & \VBC\ & 89 & 44 & 14997 & 885 & - & 95.26 & 90.86 & - & - & - & - & - \\
     &  & Baseline-BiLSTM & 76 & 810 & 14231 & 898 & - & 52.58 & 66.97 & - & - & - & - & - \\
     &  & Baseline-TextCNN & 58 & 527 & 14514 & 916 & - & 63.48 & 75.80 & - & - & - & - & - \\
    \midrule
    Draper & MCC & Russell et al.~\cite{Russell2018Draper}& - & - & - & - & - & - & 56.60 & - & 51.80 & 90.40 & 53.60 & - \\
     &  & \VBM\ & 4693 & 4598 & 114568 & 3555 & - & - & 43.34 & - & 36.24 & 87.52 & 39.44 & - \\
     &  & \VBC\ & 2168 & 6675 & 112491 & 6085 & - & - & 57.92 & - & 56.72 & 92.11 & \textbf{55.86} & - \\
     &  & Baseline-BiLSTM & 1497 & 13836 & 105330 & 6756 & - & - & 46.84 & - & 46.75 & 91.35 & 46.97 & - \\
     &  & Baseline-TextCNN & 1522 & 12266 & 106900 & 6731 & - & - & 49.40 & - & 49.02 & 91.83 & 49.24 & - \\
    \midrule
    REVEAL & F1 & REVEAL~\cite{chakraborty2020reveal}& - & - & - & - & 84.37 & 30.91 & 41.25 & 60.91 & - & - & - & - \\
     &  & \VBM\ & 84 & 269 & 1775 & 146 & 84.48 & 35.18 & \textbf{45.27} & 63.48 & - & - & - & - \\
     &  & \VBC\ & 59 & 402 & 1642 & 171 & 79.73 & 29.84 & 42.59 & 74.35 & - & - & - & - \\
     &  & Baseline-BiLSTM & 63 & 457 & 1587 & 167 & 77.13 & 26.76 & 39.11 & 72.61 & - & - & - & - \\
     &  & Baseline-TextCNN & 48 & 561 & 1483 & 182 & 73.22 & 24.50 & 37.41 & 79.13 & - & - & - & - \\
    \midrule
    muVuldeepecker & Weighted & muVuldeepecker~\cite{zou2019muvuldeepecker}& - & - & - & - & - & - & - & - & - & - & - & 96.28 \\
     & F1  & \VBM\ & 94 & 48 & 8604 & 27583 & - & - & - & - & - & - & - & \textbf{99.59} \\
     &  & \VBC\ & 94 & 65 & 8587 & 27583 & - & - & - & - & - & - & - & 99.56 \\
     &  & Baseline-BiLSTM & 8 & 8642 & 10 & 27668 & - & - & - & - & - & - & - & 65.93 \\
     &  & Baseline-TextCNN & 17 & 8639 & 13 & 27660 & - & - & - & - & - & - & - & 65.95 \\
    \bottomrule
    \end{tabular}
\end{table*}

\subsubsection{Vuldeepecker} 
The precision score reported by Vuldeepecker is 91.9\%. However, using \VBM,\ we achieved a precision score of 95.76\%, which beats the original score by 3.86\%. On top of that, our model obtained a higher F1 score, 93.03\%, compared to the score obtained by Vuldeepecker, 92.9\%, which shows that our model is balanced when classifying vulnerable and non-vulnerable samples. The low rate of false positives (0.39\%) and negatives (9.14\%) indicates that \VBM\ detects vulnerabilities both in synthetic and real-world code with a low misclassification rate. 
%
    
\subsubsection{Draper} The authors of~\cite{Russell2018Draper} use several evaluation metrics to evaluate their model on the Draper dataset. We choose MCC as a basis for comparison because it is suitable for imbalanced datasets, and the the vulnerable and non-vulnerable classes of the Draper dataset are imbalanced. \VBC\ obtained an MCC score of 55.86\% on this dataset. That is a 2.26\% increase on the performance reported in~\cite{Russell2018Draper}, and is significant as we can provide better detection while maintaining the class balance across prediction.
%

\subsubsection{REVEAL} 
The \VBM\ model achieved a higher F1 score of 45.27\% than the one of 41.25\% reported in \cite{chakraborty2020reveal}, despite not using the data rebalancing techniques proposed there.
Instead, we assigned weights to each class (vulnerable and non-vulnerable) during fine-tuning to reduce the class imbalance problem without altering the dataset. Our approach also obtained a true positive rate (TPR) higher by 2.57\%. This indicates that \VBM\ is more effective in detecting vulnerable samples correctly while also maintaining the balance with the non-vulnerable class. 
%

\subsubsection{muVuldeepecker} Differently from the previous experiments, this is a multi-class classification task. The muVuldeepecker dataset contains separate classes for 40 CWEs, so each positive sample can be mapped to a specific security vulnerability. \VBM\ achieved a very high weighted F1 score of 99.59\% compared to the 96.28\% reported in~\cite{zou2019muvuldeepecker}. It also reduced the false negatives rate (FNR) from 5.53\% to 0.41\%, which is significant as the additional samples detected as vulnerable need to be assigned to the correct class. Furthermore, looking into the prediction for specific classes such as CWE-190 and CWE-191 (integer overflow and underflow), we can see that \VBM\ is able to correctly assign more than 90\% of the vulnerable samples in their respective classes. 
%

Our models manage to surpass the performance of the compared models despite using the latter's datasets and preferred evaluation metrics. This happens across a variety of datasets with synthetic and real world data, balanced and imbalanced classes, binary and multi-class classification tasks.

\subsection{Evaluation on benchmarks}\label{sec:benchs}
We also evaluated \VBM\ and \VBC\  on two publicly available benchmarks which are used by the community as a basis for comparing different source code models on standardised tasks. In both cases, the PEM for the vulnerability detection task is Accuracy.

\subsubsection{CodeXGLUE} The CodeXGLUE~\cite{Lu2021codexglue} benchmark is the first and most popular benchmark for programming language understanding, introduced by Microsoft Research. 
We focus on the \emph{defect detection} task, which consists of vulnerability detection on the Devign dataset, described in Section~\ref{dataset:devign}. 
Table~\ref{tab:benchmark}-A shows the standings of the CodeXGLUE benchmark leaderboard at the time of publication, including our results\footnote{Leaderboard accessible at~\textit{\url{https://microsoft.github.io/CodeXGLUE}}.}. \VBM\ achieved an accuracy of 64.75\%, ranking 3rd below CoText and C-BERT. Note that \VBM\ has a significantly lower number of model parameters (55.07\%) than CoTexT, and is trained on a fraction of the data of the top 2 models. Impressively, \VBC\ achieves a slightly lower performance yet with a model size less than 1\% of CoTexT and 2\% of C-BERT.
    
\begin{table}[t]
    \scriptsize
    \caption{Benchmark evaluation results}\vspace{-5pt}
    \label{tab:benchmark}
    \centering
    \begin{tabular}{p{0.3cm} p{1.8cm} p{2.2cm} p{1.2cm} p{0.8cm}}
    \toprule
    \textbf{Rank} & \textbf{Model} & \textbf{\mbox{Pre-training data} size (\# of functions)} & \textbf{Model size} & \textbf{Accuracy (\%)}\\
    \midrule
    \multicolumn{5}{c}{\textbf{A}: CodeXGLUE leaderboard for defect detection as of 23 May 2022.}\\
    \midrule
    1 & CoTexT \cite{phan2021cotext} & 375M & 220M & 66.62\\
    2 & C-BERT \cite{buratti2020cbert} & 8.1M & 110M & 65.45\\
    3 & \textbf{\VBM}\ & \textbf{2.28M} & \textbf{125M} & \textbf{64.75}\\
    4 & \textbf{\VBC}\ & \textbf{2.28M} & \textbf{2M} & \textbf{64.42}\\
    5 & PLBART \cite{ahmad2021plbart} & 680M & 140M & 63.18\\
    6 & Code2vec \cite{coimbra202code2vec} & - & - & 62.48\\
    7 & CodeBERT \cite{Lu2021codexglue} & 8.5M & 125M & 62.08\\
    8 & RoBERTa \cite{Lu2021codexglue} & 2.4M & 125M & 61.05\\
    9 & TextCNN & - & - & 60.69\\
    10 & BiLSTM & - & - & 59.37\\
    \midrule
    \multicolumn{5}{c}{\textbf{B}: D2A leaderboard for the ``function'' category as of 23 May 2022.}\\
    \midrule
    1 & \textbf{\VBM}\ & \textbf{2.28M} & \textbf{125M} & \textbf{62.30}\\
    2 & \textbf{\VBC}\ & \textbf{2.28M} & \textbf{2M} & \textbf{60.68}\\
    3 & C-BERT \cite{buratti2020cbert} & 8.1M & 110M & 60.20\\
    4 & AugSA-S \cite{9402126d2a} & - & - & 55.20\\
    5 & AugSA-V \cite{9402126d2a} & - & - & 45.60\\
    \bottomrule
    \end{tabular}
\end{table}

\subsubsection{D2A} D2A~\cite{9402126d2a} is a benchmark for vulnerability detection introduced by IBM Research. It is based on the D2A dataset, described in Section~\ref{dataset:d2a}. Table~\ref{tab:benchmark}-B shows the standings of the D2A benchmark leaderboard at the time of publication, including our models\footnote{Leaderboard  accessible at \textit{\url{https://ibm.github.io/D2A}}.}. \VBM\ tops the leaderboard for the ``function'' task (the only one relevant to our approach) with an accuracy of 62.30\%, and \VBC\ is second with 60.68\%. Interestingly, on this dataset both our models outperform C-BERT, despite its larger pre-training dataset, and the larger number of parameters (in comparison to \VBC). 
    
    
\subsection{Discussion}
Table \ref{tab:finetuning_results} shows that \VBM\ has the best PEM score on 3 out of 4 datasets and \VBC\ has the best PEM score on the remaining dataset. However, if we compare our two models on the full spectrum of fine-tuning results, we notice that the difference between the two is negligible. 
Similarly, we can see a close relationship between \VBM\ and \VBC\ on the benchmark evaluation. Both models consistently ranked next to each other and achieved state of the art performance across both benchmarks.
%
%
A possible explanation is that the code representation knowledge inherited from the pre-trained \VB\ model plays a crucial role for vulnerability detection. 

%
Model size and pre-training data size commonly play an essential role in learning better code representations. However, VulBERTa contains only 125M parameters, and our pre-training data contains only 2.28M C/C++ functions compared, for example, to CoTexT (375M) and C-Bert (8.5M). We are on par with these approaches by delivering SOTA detection performance with smaller models and pre-training data.

Based on our preliminary analysis of testing and implementing different tokenisation techniques, we believe that our tokenisation approach plays an important role in making syntactic and semantic information easily available to the simple neural architectures that use it for fine tuning.

In fact, model simplicity seems to be part of the solution. 
\VBC, our simple TextCNN-based approach (with only 2M parameters) has an MCC score of 55.86\% on the Draper dataset, higher than the 52\% obtained by the complex 3GNN model recently proposed in~\cite{zhuang2021softwareibm}, which uses a Crystal Graph Convolutional Network and Self Attention Pooling.


\subsection{Limitations}
Despite the encouraging results, we identify a number of limitations in our approach.
A common and still unsolved problem with vulnerability detection datasets is that manual inspection reveals occasional label inaccuracies. While deep learning should be resilient to label noise in training, the presence of noise during testing somewhat undermines the quantitative preformance results. %
Although our models are relatively small, they are still expensive to train. Due to limited resources, we could not explore significantly larger model configurations and combinations, including performing hyper-parameter sweep. Therefore it is possible that different \VB\ configurations could achieve a higher performance.
Finally, the main limitation of our work is the lack of a systematic attempt to detect novel 0-days vulnerabilities in open-source projects \emph{in-the-wild}. This was due to the challenge of manually reviewing false positives, which 
we would like to address in future work, leveraging explainability techniques.

\section{Conclusions}
We have presented \VB, a pre-training model which learns a deep representation of C/C++ code.
A key component of our model is a custom tokenisation pipeline which aims to preserve syntactic and semantic knowledge from source code without recurring to overly complicated neural architectures. 
We used the \VB\ model as the basis to fine-tune two models for vulnerability detection which achieve state of the art performance on several datasets and benchmarks.
This models stand out in virtue of their conceptual simplicity and low complexity as measured in terms of number of parameters.
 
 \noindent\hspace{5pt}\textbf{Acknowledgments.} This work was partially supported by the Google Cloud Research Credits program with the award GCP19980904.


\bibliographystyle{IEEEtran}
\bibliography{main_ieee}

\end{document}